\documentclass[a4paper]{elsarticle}

\usepackage[english]{babel}
\usepackage[utf8x]{inputenc}
\usepackage[T1]{fontenc}

\usepackage[a4paper,top=3cm,bottom=2cm,left=3cm,right=3cm,marginparwidth=1.75cm]{geometry}

\usepackage{amsmath}
\usepackage{graphicx}
\usepackage[colorinlistoftodos]{todonotes}
\usepackage[colorlinks=true, allcolors=blue]{hyperref}
\usepackage{subcaption}
\usepackage{float}

\title{\textbf{The role of slip transfer at grain boundaries in the propagation of microstructurally short fatigue cracks in Ni-based superalloys}}

\author{M. Jiménez$^{a,b}$}
\author{W. Ludwig$^{c,d}$}
\author{D. González$^{e}$}
\author{J.M. Molina-Aldareguia$^{a,*}$}

\address{$^{a}$IMDEA Materials Institute, C/Eric Kandel 2, 28906 Getafe, Madrid, Spain \\ $^{b}$Dpt.of Materials Science and Engineering, Universidad Carlos III de Madrid (UC3M), 28911 Leganés, Spain \\$^{c}$Université de Lyon, INSA Lyon, Mateis, UMR CNRS 5510, Lyon, France \\ $^{d}$European Synchrotron Radiation Facility, 38043 Grenoble, France \\ $^{d}$PRISM$^{2}$ Research Group, School of Metallurgy  Materials,
The University of Birmingham, Edgbaston, Birmingham, B15 2TT, UK. \\ *Corresponding author. E-mail: jon.molina@imdea.org}

\begin{document}

\begin{abstract}

Crack initiation and propagation under high-cycle fatigue conditions have been investigated for a polycrystalline Ni-based superalloy by \textit{in-situ} synchrotron assisted diffraction and phase contrast tomography. The cracks nucleated along the longest coherent twin boundaries pre-existing on the specimen surface, that were well oriented for slip and that presented a large elastic incompatibility across them. Moreover, the propagation of microstructurally short cracks was found to be determined by the easy slip transfer paths across the pre-existing grain boundaries. This information can only be obtained by characterization techniques like the ones presented here that provide the full set of 3D microstructural information. \\

 
\end{abstract}

\maketitle

Ni-based superalloys have been a critical material for jet turbine engines for decades, and thus, their fatigue crack propagation behavior has been extensively studied using linear elastic fracture mechanics (LEFM). However, in most applications, the material is subjected to high-cycle fatigue (HCF), and thus, a large fraction of the fatigue life is consumed in the initiation and propagation of microstructurally short cracks. The understanding of their mechanisms remains a major challenge, because both processes are very sensitive to the local microstructure. For instance, fatigue damage initiation involves localized plastic deformation in individual grains, in the form of persistent slip bands (PSBs)\cite{mughrabi1983fatigue}. In Ni-based superalloys, with an FCC lattice, dislocation glide takes place in the preferred $\left\lbrace111\right\rbrace$<1$\overline{1}$0>  slip system, and thus, damage typically initiates along $\left\lbrace111\right\rbrace$ planes. Moreover, they are prone to the formation of $\left\lbrace111\right\rbrace$<11$\overline{2}$> annealing twins during processing \cite{xia2006effects} and coherent twin boundaries (CTB) have been reported to have a large influence on fatigue crack initiation \cite{stinville2016combined}. In addition, the grain size and the presence of non-metallic inclusions have also been reported as important microstructural parameters for crack nucleation \cite{larrouy2015grain,texier2016crack,deng2015grain,abuzaid2013plastic}. 

\begin{figure}[h]
   \centering
    \begin{subfigure}[b]{0.5\textwidth}
        \includegraphics[width=\textwidth]{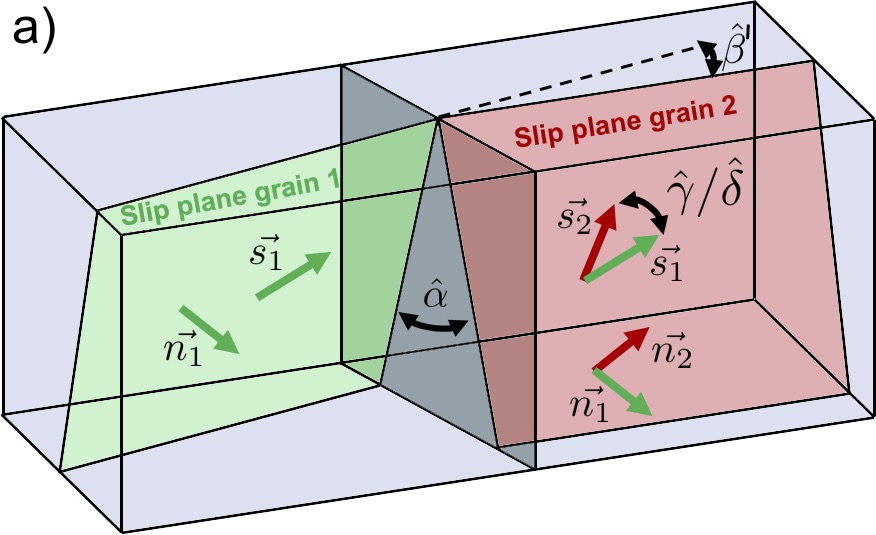} 
    \end{subfigure}
    
    \vspace{0.5 cm}
    \begin{subfigure}[b]{0.5\textwidth}
        \includegraphics[width=\textwidth]{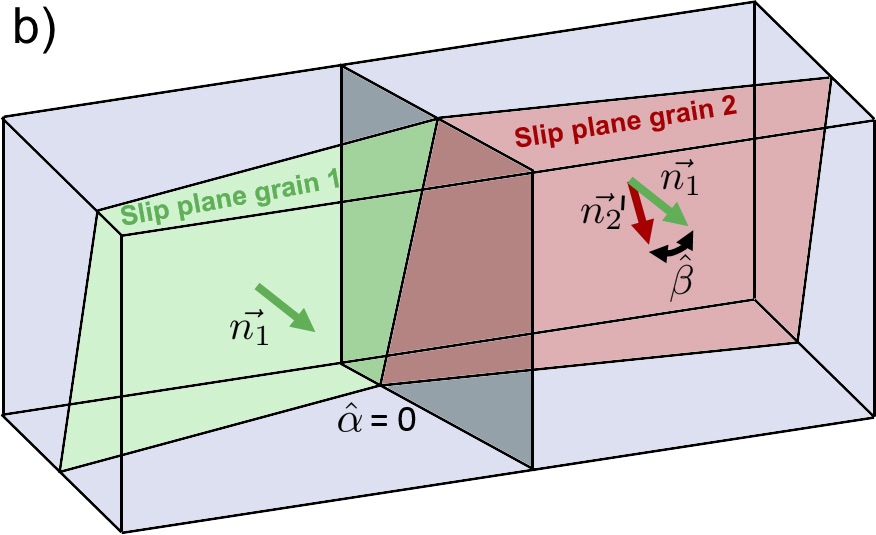}
    \end{subfigure}
    \caption{Definition of a) twist angle $\hat{\alpha}$ (angle between the traces of the slip/crack planes on the GB), $\hat{\gamma}$ angle (angle between the incoming slip direction and the direction with maximum SF of the outgoing plane) and $\hat{\delta}$ angle (minimum angle between the incoming slip direction and outgoing slip directions for each possible outgoing plane). b) tilt angle $\hat{\beta}$ (angle between slip/crack planes normals after rotating the outgoing slip plane around the vector normal to the GB until the twist angle $\hat{\alpha}$ is reduced to 0). }
    \label{fig:Angles}  
    
\end{figure}

Regarding the propagation of microstructurally short cracks, the first stages should be controlled by the interaction of the PSBs, with the neighboring grain boundaries (GB). A stage-I fatigue crack propagates along a single slip plane within an individual grain \cite{boyd1994short,liu2011situ}. Thus, when it reaches a GB, it must change its path to the favorable slip plane of the outgoing grain. It was early on addressed that the misorientation between the incoming and outgoing slip/crack planes, rather than the lattice misorientation between the two grains, should be the key factor that controls crack propagation across the GB. Previous studies \cite{kumar2007crack,reed2009fatigue,zhai2000crystallographic,holzapfel2007interaction} defined the misorientation between the two crack planes by the twist and tilt components along the GB ($\hat{\alpha}$ and $\hat{\beta'}$, respectively, in figure \ref{fig:Angles}a, where the twist was measured as the angle between the intersections of the two crack planes with the GB plane and the tilt, as the angle between the two crack traces at the observation surface. However, most previous studies are limited because they were conducted in \textit{post-mortem} specimens \cite{holzapfel2007interaction} and/or rely on two-dimensional inspection techniques, like scanning electron microscopy (SEM) and electron backscatter diffraction (EBSD) analysis \cite{larrouy2015grain,texier2016crack,stinville2016combined,boyd1994short,kumar2007crack,reed2009fatigue,ma2010situ}. In such investigations, the GB plane orientation below the surface is unknown and therefore is ignored, so that the quantification of the tilt and twist angles is arbitrary and dependent on the observation surface. Moreover, even though it is clear that the initial stages of crack propagation are driven by plasticity \cite{kumar2007crack,kim2013situ}, slip transfer and thus crack propagation at the GB should not only be controlled by the misorientation between the incoming and outgoing \textit{planes}. The misorientation between the preferred slip directions of the incoming and outgoing grains should also play a key role in the propagation of cracks and this parameter has been systematically ignored by previous analyses due to the limitations of the conventional inspection techniques. Therefore, in order to provide the material design with critical data to improve fatigue life in polycristalline Ni-based superalloys, there is a requirement for an \textit{in-situ }technique that can address the role of the three-dimensional microstructure on the early stages of fatigue crack initiation and propagation. As first demonstrated in \cite{Herbig:2011, King:2011a}, a combination of state-of-the-art synchrotron X-ray diffraction and imaging techniques fulfills this requirement. Phase contrast tomography (PCT) exploits the coherence of synchrotron beams and enables non-destructive, time-lapse observation of fatigue crack propagation in miniaturized fatigue specimens with a resolution of 10 µm \cite{cpb97}, even if their opening displacement is still below the spatial resolution of the X-ray imaging system.  X-ray diffraction contrast tomography (DCT) on the other hand is a full-field diffraction imaging technique exploiting Bragg diffraction enabling the reconstruction of spatially resolved 3D orientation maps for a large variety of polycrystalline materials \cite{Ludwig09b, Reischig:2013a, Vigano2016}.

In this work, PCT and DCT were carried out \textit{in-situ} during fatigue testing in a forged polycrystalline Inconel 718 (IN718) alloy at the ID11 beamline of the European Synchrotron Radiation Facility (ESRF). IN718 is extensively used in the low-pressure turbine section of jet engines due to its excellent mechanical strength and fatigue resistance at moderate temperatures. It presents a very complex microstructure, with randomly oriented equiaxed  $\gamma$-Ni grains, with an average grain size of 100 $\mu$m, and a large volume fraction of annealing twins. Its strength relies on solid-solution and precipitation strengthening by $\gamma$' and $\gamma$'' phases. It also shows the presence of other micrometer size second-phase particles, including carbides (with an average size of 2$\mu$m) and $\delta$-phase particles. The fatigue specimens had a gauge length of 1 mm and rounded cross sectional area of 0.36 mm$^{2}$ (approximately 600$\mu$m on each side), optimized for this type of combined DCT and PCT analysis and \textit{in-situ }testing inside the Nanox fatigue rig (more details about the in-situ testing rig can be found in \cite {Gueninchault2016}). The specimen was manufactured by electro discharge machining and subsequent rotative electropolishing to reduce stress concentrations associated to sharp corners and surface roughness, ensuring that crack initiation was microstructure (and not geometry) driven while removing the surface carbides. Three DCT measurements covering a height of 600 µm were acquired prior to the test to determine the initial 3-D grain structure, while PCT was carried out at regular intervals of 20,000 cycles to track the initiation and propagation of the cracks until final fracture. The test was conducted imposing a constant strain amplitude resulting in a maximum stress of 700 MPa (around 2/3 of the yield stress), stress ratio R = $\sigma_{min}/\sigma_{max}$ = 0.1 and a frequency of 25 Hz . During the test, 3 cracks nucleated simultaneously at the surface before 40,000 cycles. Initially, the crack fronts were parallel to the surface, but they quickly evolved into a semi elliptical shape. Two of the cracks merged into a single one that eventually led to fracture after 108,000 cycles.

\begin{figure}[h]
    \centering
    \begin{subfigure}[b]{0.48\textwidth}
        \includegraphics[width=\textwidth]{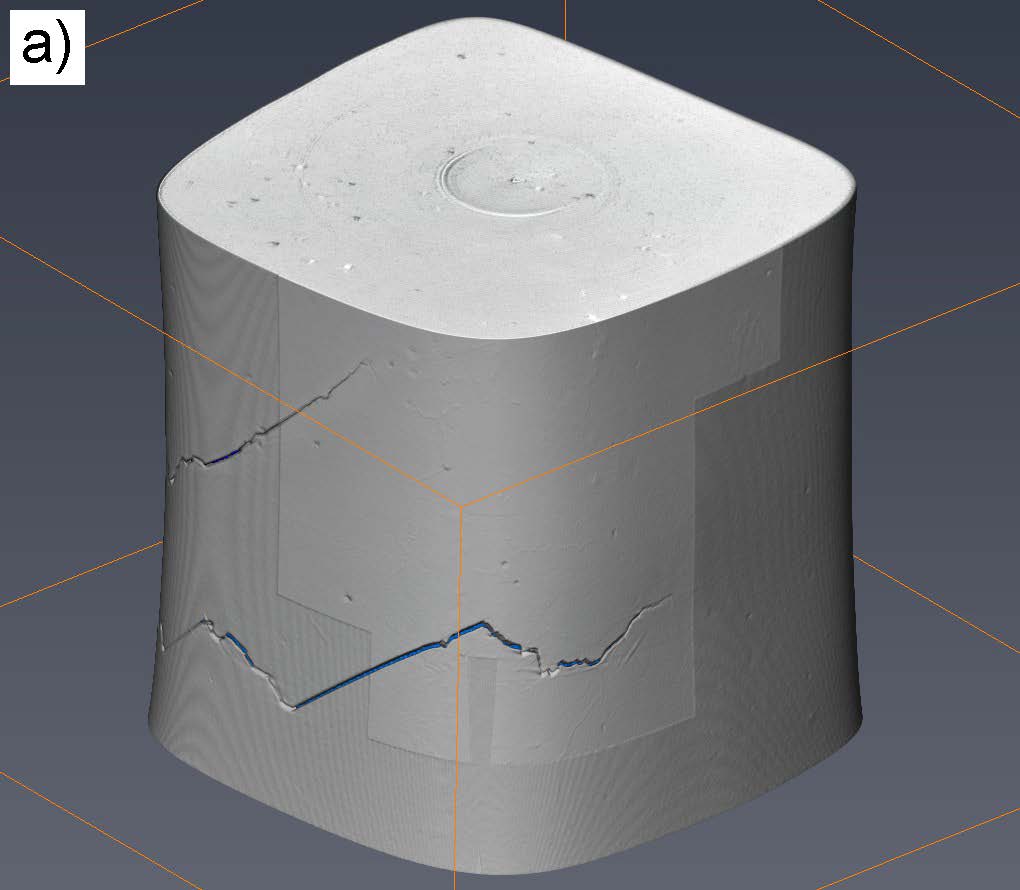}
    \end{subfigure}
    \hspace{0.125 cm}
    \vspace{0.4 cm}
    \begin{subfigure}[b]{0.48\textwidth}
        \includegraphics[width=\textwidth]{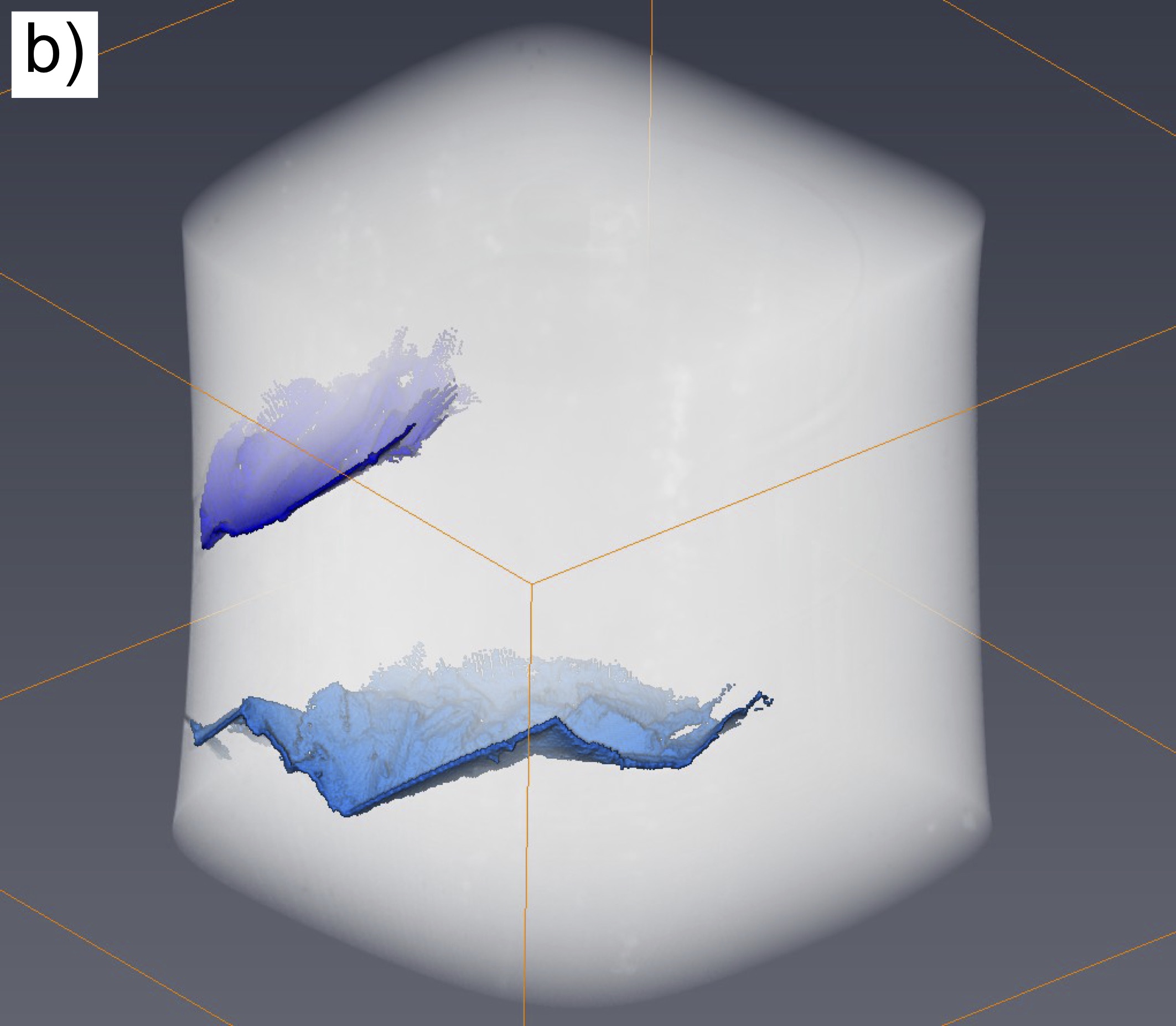}
        \end{subfigure}
    \begin{subfigure}[b]{0.48\textwidth}
        \includegraphics[width=\textwidth]{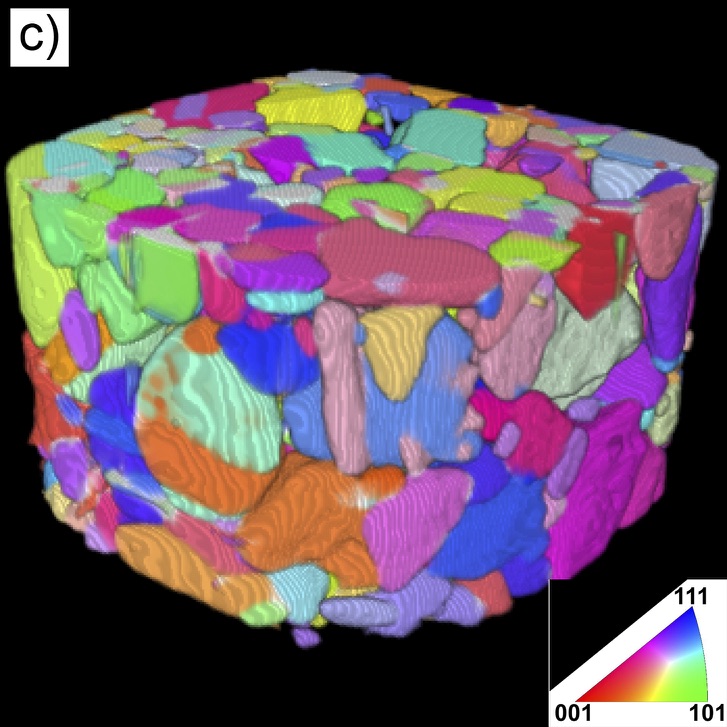}
        \end{subfigure}
    \hspace{0.125 cm}
    \begin{subfigure}[b]{0.48\textwidth}
        \includegraphics[width=\textwidth]{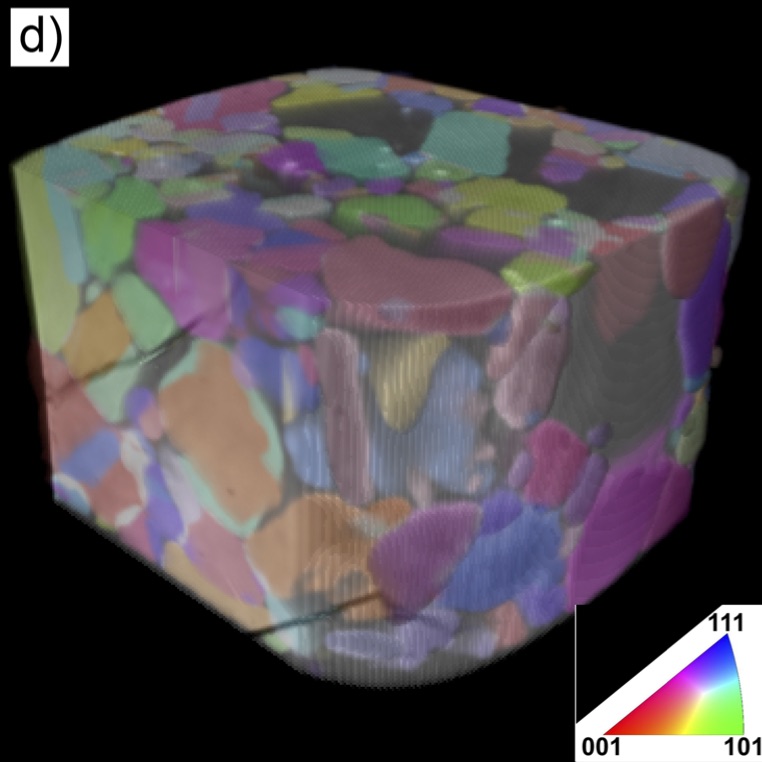}
    \end{subfigure}
    
    \caption{Reconstructed volumes of: a) PCT after 100,000 cycles. Cracks can be seen on the surface. b) PCT after 100,000 cycles with segmented cracks. c) DCT. d) Interior slice of the DCT overlapped with the PCT after 100,000 cycles, showing the interaction of the cracks with the specimen microstructure in 3D. Orientations are referred to the load (vertical) axis. The side length of the specimen is 600$\mu$m.}
    \label{fig:volumes}
\end{figure}

The PCT volumes showed high quality reconstructions and clear images of the cracks (figures \ref{fig:volumes}a and \ref{fig:volumes}b). The DCT volume is shown in figure \ref{fig:volumes}c. About 80 \% of the sample volume could be indexed and reconstructed. The indexing and reconstruction process is affected by the presence of extended orientation related domains (c.f. large fraction of Sigma 3 boundaries), which give rise to systematic diffraction spot overlaps. Figure \ref{fig:volumes}d shows the overlay of both, the PCT and DCT volumes. Cracks could be clearly tracked and confronted with the local microstructure, allowing the identification of the individual grains that led to crack initiation, the crystallographic planes along which the cracks propagated and the change in crack plane when the cracks were transmitted across a GB.

\begin{figure}[h]
    \centering
    \begin{subfigure}[b]{0.47\textwidth}
        \includegraphics[width=\textwidth]{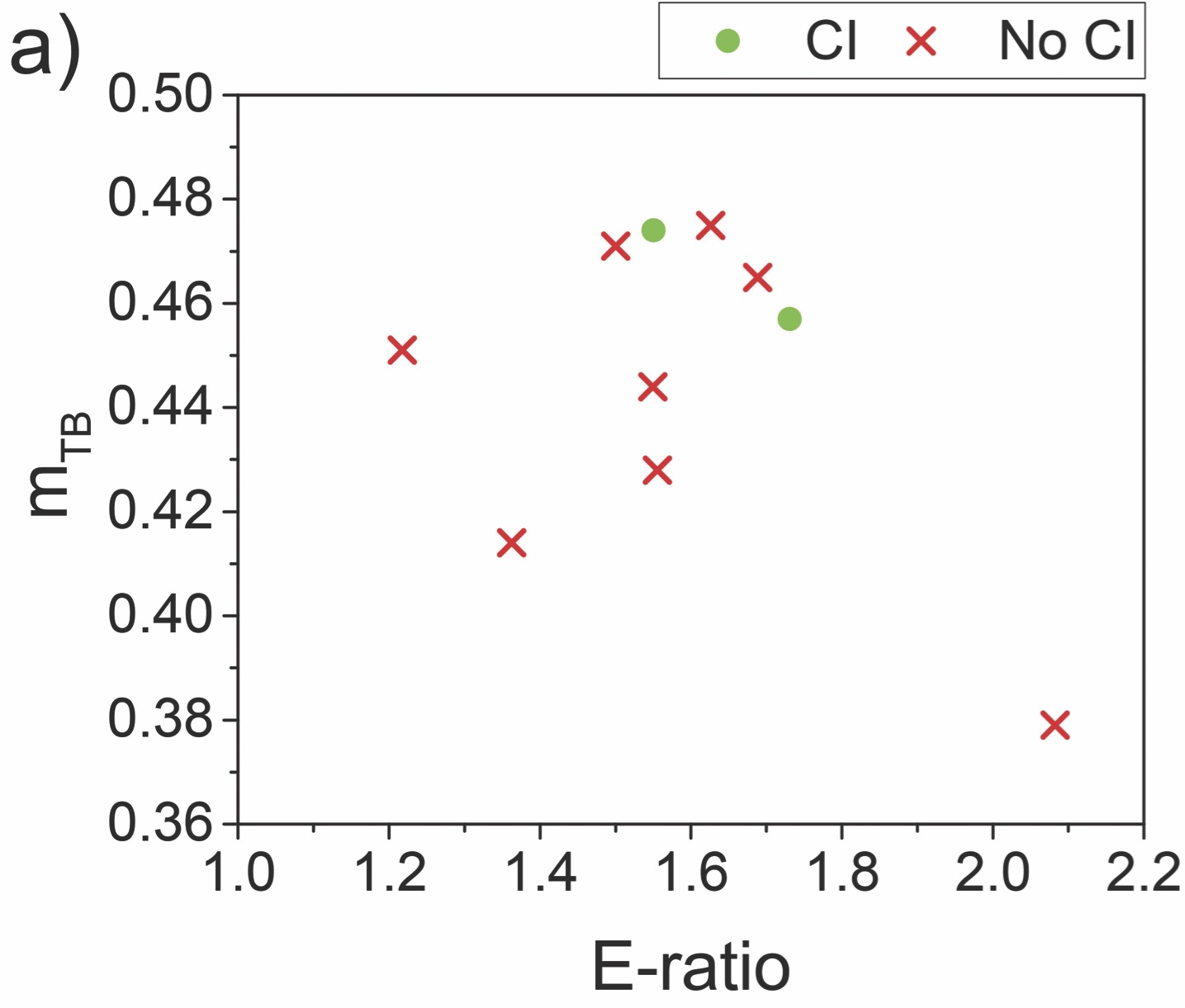} 
    \end{subfigure}
    \begin{subfigure}[b]{0.48\textwidth}
        \includegraphics[width=\textwidth]{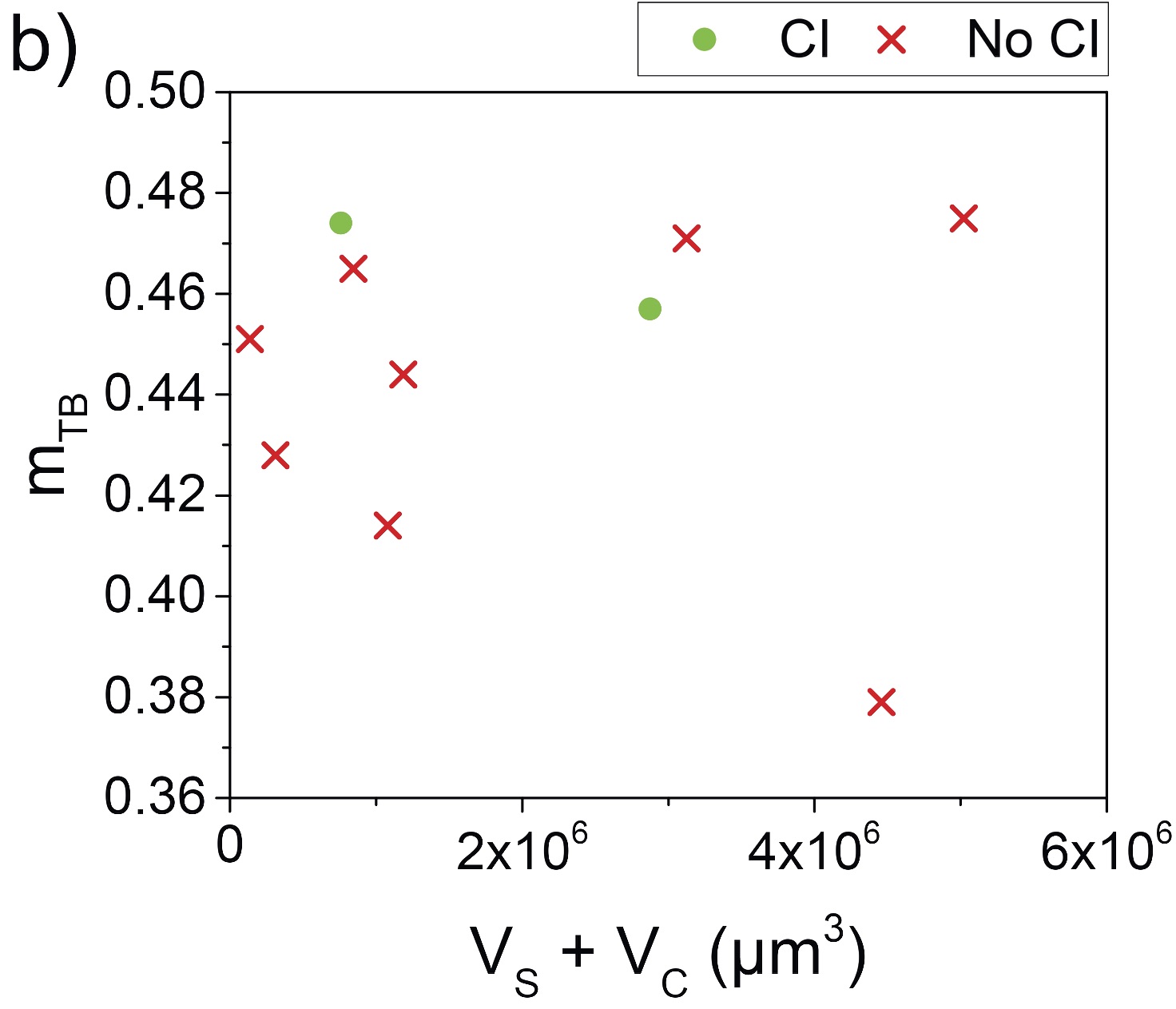}
    \end{subfigure}
    
    \vspace{0.5 cm}
    \begin{subfigure}[b]{0.5\textwidth}
        \includegraphics[width=\textwidth]{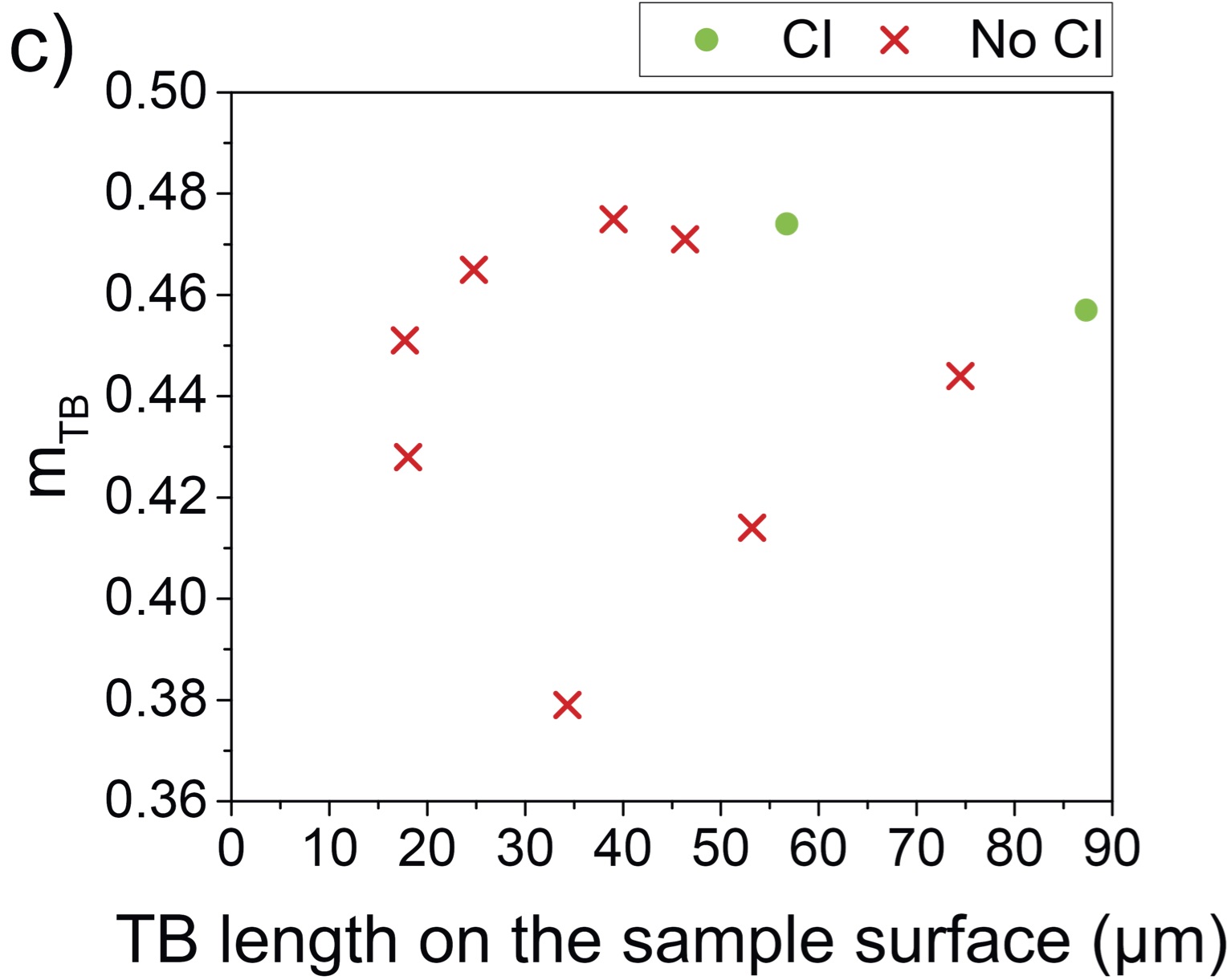}
    \end{subfigure}
    \vspace{0.5 cm}
    \begin{subfigure}[b]{0.46\textwidth}
        \includegraphics[width=\textwidth]{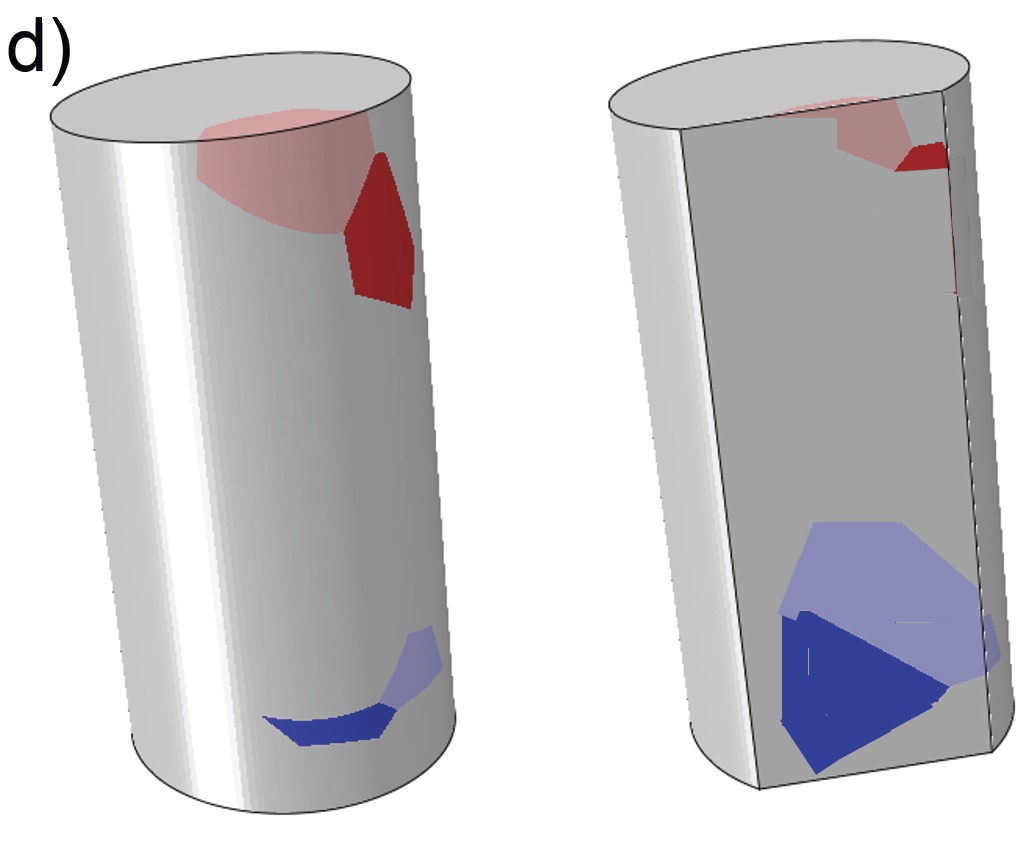}
    \end{subfigure}

    \caption{The applied Schmidt factor on all pre-existing CTBs found at the specimen surface is plot against different factors that can potentially lead to fatigue crack initiation. Green circles represent those CTBs that led to fatigue crack initiation (CI), while the red crosses stand for the ones that did not (No CI). a) E-ratio = E$_{1}$/E$_{2}$. b) Sum of volumes of the grains at both sides of the TB. c) Length of the TB along the specimen surface. d) Schematic illustration of the influence of the volume of grains and length of the CTB on the surface. A long CTB on the surface that separates a small pair of grains (red grains) is more prone to crack initiation than a shorter CTB on the surface that separates two bigger grains (blue grains).}
    \label{fig:CI}
\end{figure}

Two out of the three initial cracks were located in the specimen gauge volume covered by DCT. Both cracks nucleated on the specimen surface along a slip system parallel to a preexisting CTB. The results indicate that preexisting CTBs are more relevant than the second phase particles in fatigue crack initiation in this case. However, a large number of CTBs were present on the specimen surface, and not all of them led to cracking. In order to rationalize which CTBs led to crack initiation, the following factors were analyzed: the applied Schmidt factor on the slip system parallel to the CTB, m$_{TB}$, the elastic contrast across the CTB (measured as the ratio between the elastic moduli along the loading axis, E-ratio = E$_{1}$/E$_{2}$, of the two grains at each side of the TB), the grain size (measured as the sum of the volume of the two grains,V$_{1}$+V$_{2}$) and the length of the CTB on the specimen surface. Figure \ref{fig:CI} summarizes the results, where the green circles represent those CTBs that led to crack initiation and the red crosses those that did not. It is clear that for the cases where crack initiation occurred, a large applied Schmidt factor and a large E-ratio of 1.7 was found (figure \ref{fig:CI}a), implying that in those grains well oriented for slip a large elastic incompatibility across the CTB can explain the local stress concentration that leads to damage nucleation. However, large E-ratios and Schmidt factors are two necessary, but not sufficient conditions for the nucleation of cracks, as for instance a pre-existing CTB with a E-ratio=2.1 did not lead to crack initiation. It is also expected that larger grains will be softer and therefore more prone to damage than small ones, but figure \ref{fig:CI}b shows that this is not the most important factor because the largest grains did not lead to crack initiation. On the contrary, figure \ref{fig:CI}c shows that the length of the CTB, rather than the volume of the grains, is the most relevant characteristic size leading to crack initiation: the longer the intersection of the TB with the surface, the more likely it is to initiate a crack. This is explained by the more relaxed constraint of the slip plane parallel to the CTB on the surface, which facilitates the formation of the PSBs. As shown schematically in figure \ref{fig:CI}d, larger grains for which the intersection of the CTB with the surface is short (for instance, the blue pair of grains) are less susceptible to initiate cracks than smaller grains that display a larger CTB on the surface (the red pair of grains). Even though the current observations in a single specimen are limited, the conclusions are in full agreement with those reported using conventional testing \cite{stinville2016combined}, but with the advantage that the tomography results allow measuring the sub-surface shape and size of the grains and orientation of the nucleated cracks, which is impossible with surface inspection techniques.    

Regarding crack propagation across GBs, contrary to previous surface observations \cite{zhai2000crystallographic}, the current measurements provide full access to all relevant geometrical parameters that define the propagation of cracks across each GB, including the real GB orientation below the surface: the real twist angle, in figure \ref{fig:Angles}a, is measured as the angle between the intersection of the two slip/crack planes with the GB plane; and the real tilt angle $\hat{\beta}$, in figure \ref{fig:Angles}b, as the angle between the two slip/crack plane normals, once the outgoing slip/crack plane is rotated around the GB normal to remove the twist. Moreover, since in Stage-I, the cracks propagate parallel to a slip plane in each grain, for each incoming slip/crack plane on an individual GB, there are four potential outgoing slip planes and three potential slip directions, and the experimental data can be used to evaluate the role of slip transferability at the GB in front of the crack tip on the selection of the outgoing slip/crack plane. To account for this, starting from the initial grain, two geometrical parameters were computed for each of the four potential slip/crack planes at each of the 12 GBs encountered by the crack: $\hat{\gamma}$, which is the angle between the incoming slip direction and the slip direction with the highest Schmidt factor for each potential outgoing slip/crack plane, and $\hat{\delta}$, which is the minimum angle between the slip direction of the incoming plane and the best aligned slip direction of each outgoing plane. The first one accounts for how well the outgoing slip system is oriented with respect to the externally applied load, while the second one evaluates how the geometrical alignment between the incoming and outgoing slip systems affects crack propagation across the GB. Comparison of these geometrical parameters for the actually observed propagation crack paths with respect to those potentially available allowed discriminating between the most important factors leading to crack propagation at GBs.

\begin{figure}[H]
    \raggedright    
    \begin{subfigure}[b]{0.48\textwidth}
        \includegraphics[width=\textwidth]{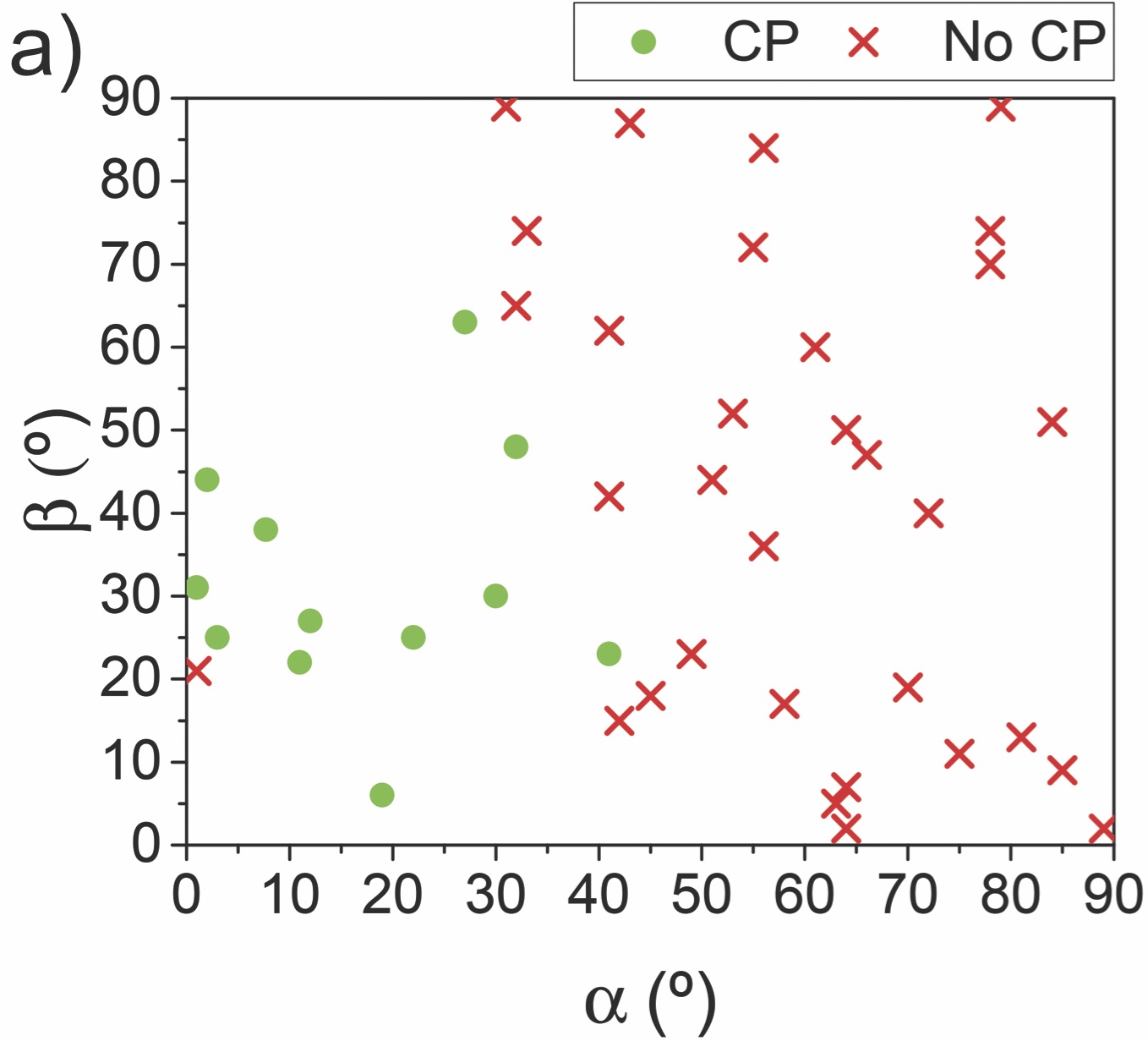}
    \end{subfigure}
    \hspace{0.3 cm}
    \begin{subfigure}[b]{0.48\textwidth}
        \includegraphics[width=\textwidth]{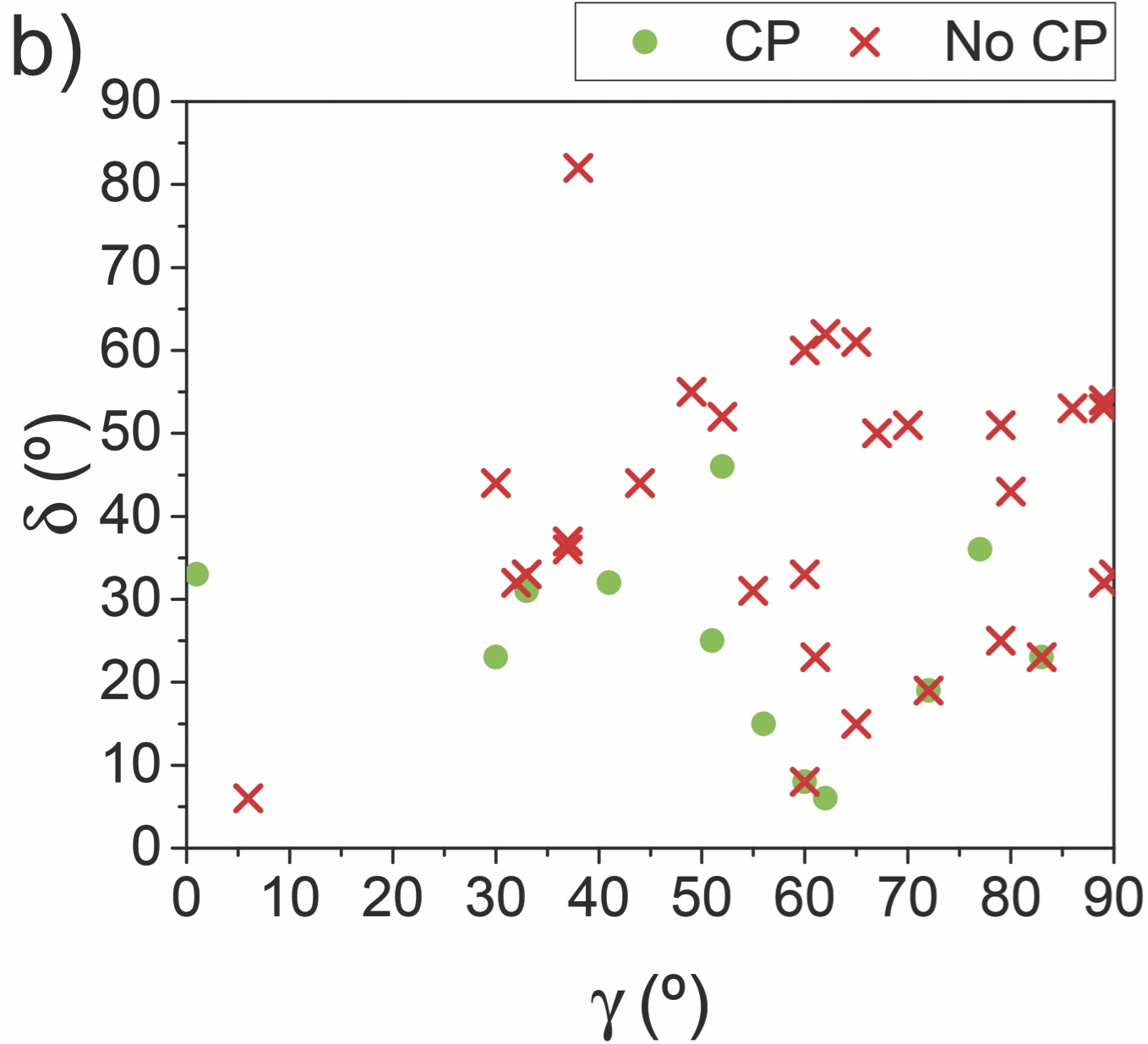}
    \end{subfigure}
    
    \vspace{0.5 cm}
    \hspace*{\fill}
    \begin{subfigure}[b]{0.44\textwidth}
        \includegraphics[width=\textwidth]{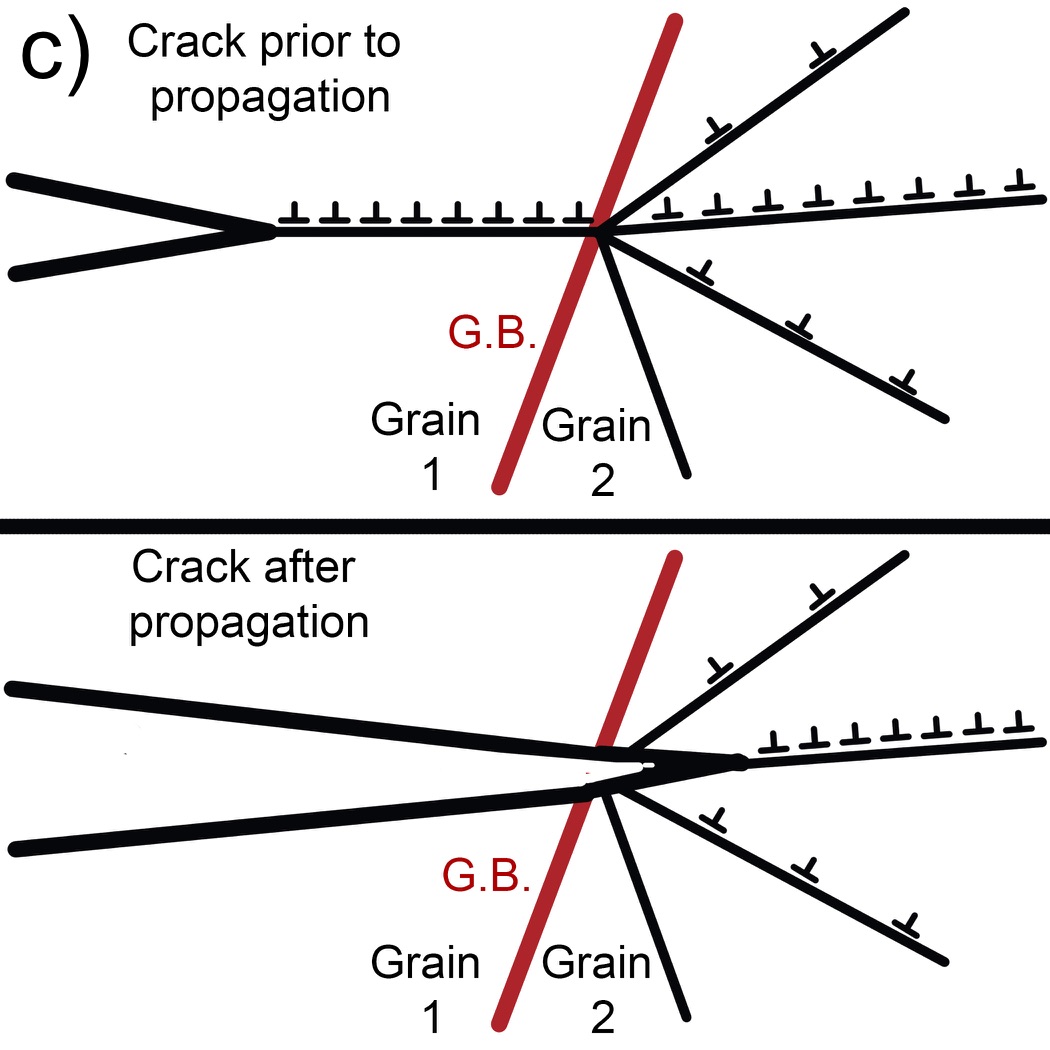}
    \end{subfigure}
    \hspace{0.65 cm}
    \begin{subfigure}[b]{0.48\textwidth}
        \includegraphics[width=\textwidth]{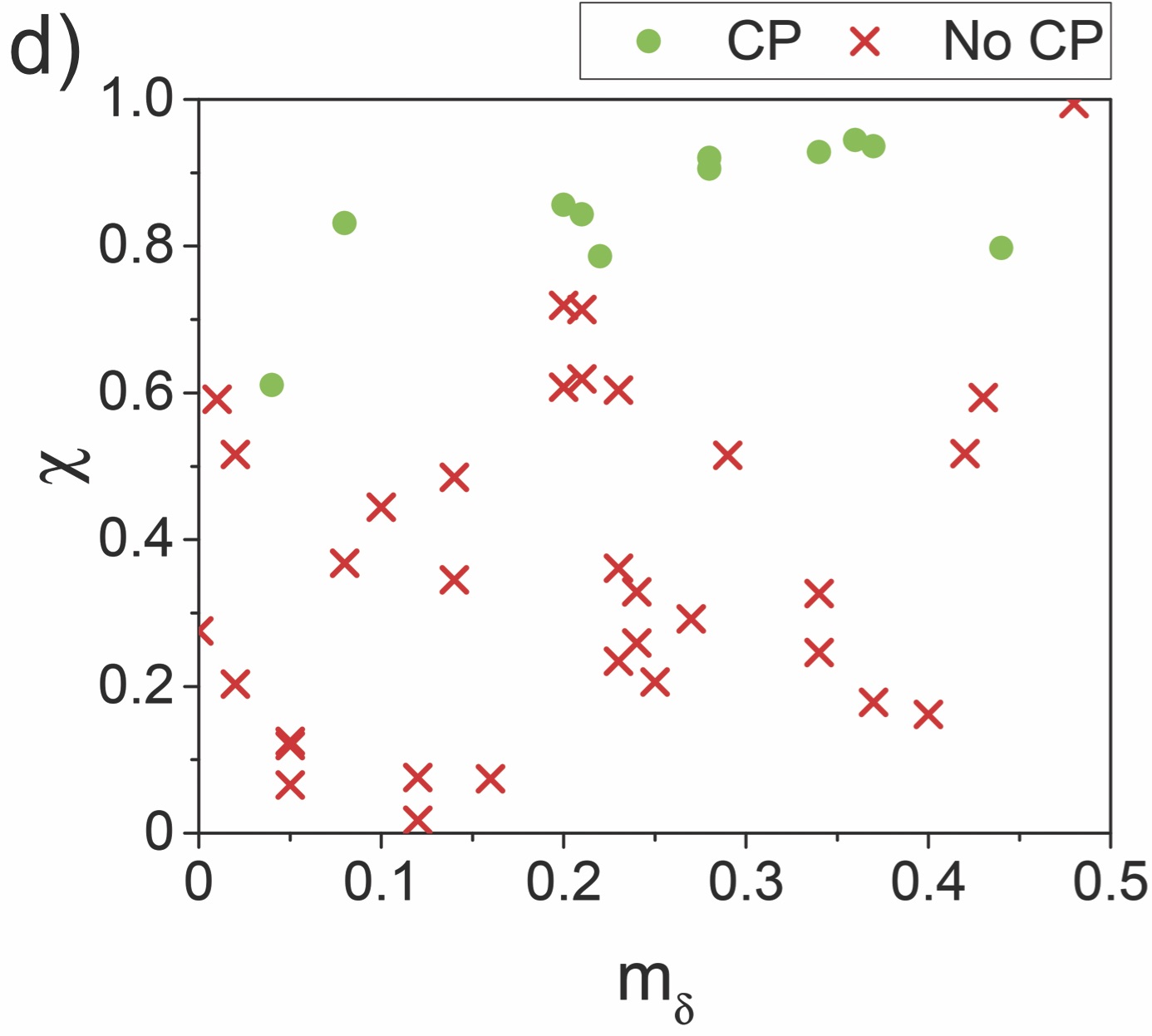}
    \end{subfigure}
    \caption{Analysis of crack propagation across GBs. Green circles represent the conditions for which crack propagation occurred (CP) and the red crosses represent other potential crack paths that were not observed (No CP). a) Values of the angles $\hat{\alpha}$ and $\hat{\beta}$. b) Values of the angles $\hat{\delta}$ and $\hat{\gamma}$. c) Schematic illustration of how slip transfer associated with a growing fatigue crack can determine the crack propagation path across a GB in Stage-I. d) Value of the ${\chi}$ factor versus the applied Schmidt factor.}
    \label{fig:CP}
\end{figure}

Figure \ref{fig:CP}a plots the $\hat{\alpha}$ and $\hat{\beta}$ angles of all potential slip/crack planes for each analyzed GB impinged by the fatigue crack, with the green full circles representing actual observed crack paths and the red crosses, potential planes at which the cracks did not propagate. It is clear that crack propagation occurred predominantly for slip planes displaying a small twist angle, $\hat{\alpha}$ < 45º, while the tilt angle, $\hat{\beta}$, did not seem to influence significantly the crack path. The predominant influence of $\hat{\alpha}$ on crack propagation across GBs is in full agreement with previous observations \cite{zhai2000crystallographic}, and it is reasonable because the lower the value of $\hat{\alpha}$, the smaller is the GB surface that the crack has to break through in order to propagate to the adjacent grain. Similarly, figure \ref{fig:CP}b plots the $\hat{\delta}$ and $\hat{\gamma}$ angles for all observed (green circles) and potential slip/crack planes (red crosses) for each analyzed GB impinged by the fatigue crack. It is clear that planes with a small $\hat{\delta}$, < 45º, were always preferred propagation paths, whereas the impact of $\hat{\gamma}$ was found negligible. Therefore, both a small twist angle, $\hat{\alpha}$, and a small angle between the incoming and outgoing slip directions, $\hat{\delta}$, are key determining the crack path across GBs. Interestingly, both are geometrical parameters associated with easy slip transmission across GBs \cite{genee2017slip}. Therefore, the results allow concluding that crack propagation in stage-I is facilitated by slip transmission, as schematically illustrated in figure \ref{fig:CP}c. In other words, the plasticity induced at the crack tip in the incoming grain, predominantly parallel to the growing fatigue crack, activates outgoing slip systems in adjacent grains that are well oriented for slip transfer, and that dictate the crack path across the GB. 

If this is so, it is illustrative to use an additional parameter to rank the potential crack paths that include both $\hat{\alpha}$ and $\hat{\delta}$ and that has been used before in the context of slip transmission at GBs \cite{beyerlein2012structure}, the factor $\chi$, defined as the product of the cosines of those angles:

\begin{equation}
\chi = cos(\hat{\alpha}) \times cos(\hat{\delta})
\end{equation}

The  $\chi$ factor, ranges between 0 and 1, with 1 implying the best possible geometrical alignment between the incoming and outgoing slip systems and 0 the worst one. Figure \ref{fig:CP}d clearly shows that the observed CP planes correspond to slip systems with a $\chi$ factor close to 1, regardless the applied Schmidt factor on the outgoing slip system, m$_{TB}$. In other words, the fatigue crack propagates along directions of easy slip transfer, irrespective of their orientation with respect to the applied load. Interestingly, it was found that in the few cases where the crack did not follow a path with a high $\chi$ factor, this was associated to either bifurcation events at a given GB, due to the existence of two slip transfer paths with a high $\chi$ factor, or to cases where the incoming crack impinged on a pre-existing CTB, in which case the crack propagated along it, until impinging in the next GB.

In conclusion, the combination of DCT and PCT has been proven a very useful tool for the study of fatigue crack nucleation and propagation. It allows \textit{in-situ} analysis of the crack path evolution with respect to the complete tridimensional microstructure. This work confirms that those long CTBs well oriented for slip pre-existing on the specimen surface are the preferred crack initiation sites. The propagation of MSC in stage-I occurs parallel to individual slip planes in each grain and is controlled by the easy slip transfer path at each GB. Easy slip transfer paths are those corresponding to small twist angles, $\hat{\alpha}$ and $\hat{\delta}$ angles. The  ${\chi}$ factor, typically used to evaluate slip transfer at GB, is proposed as the key factor determining the fatigue crack propagation path through the grain boundaries.\\ 

\textbf{Acknowledgments}\\

The authors want to acknowledge the contribution of Nicolas Gueninchault, Andrew King, Joseph Kelleher, Colin Lupton and Tim Wigger in the Synchrotron experiments. This work was supported by the State Secretariat for Research, Development and Innovation of the Spanish Ministry of Economy and Competitiveness through the project CRACKTIAL (MAT2016-77189-R).\\

\textbf{References}
\bibliographystyle{elsarticle-num}
\bibliography{bibliography.bib}

\end{document}